\documentclass[preprint,amsmath,amssymb,aps,nofootinbib,showpacs]{revtex4}
\usepackage{graphicx}
\usepackage{bm}
\usepackage{subfigure}
\usepackage{amsmath}

\newcommand{\epl}{Europhys. Lett.\ }

\newcommand{\jpa}{J. Phys. A\ }

\newcommand{\etal}{{\em et al.}}

\newcommand{\UQ}{School of Mathematics and Physics, University of Queensland, Brisbane, 
QLD 4072, Australia.}

\begin{document}
\title{Quantum phase-space analysis of population equilibration in multi-well ultracold atomic systems}

\author{C.V. Chianca and M.K. Olsen}
\affiliation{\UQ}
\date{\today}

\begin{abstract}

We examine the medium time quantum dynamics and population equilibration of two, three and four-well Bose-Hubbard models using stochastic integration in the truncated Wigner phase-space representation. We find that all three systems will enter at least a temporary state of equilibrium, with the details depending on both the classical initial conditions and the initial quantum statistics. We find that classical integrability is not necessarily a good guide as to whether equilibration will occur. We construct an effective single-particle reduced density matrix for each of the systems, using the expectation values of operator moments, and use this to calculate an effective entropy. Knowing the expected maximum values of this entropy for each system, we are able to quantify the different approaches to equilibrium. 

\end{abstract}

\pacs{03.75.Lm, 03.75.Kk, 67.25.du, 03.75.Gg}

\maketitle

\section{Introduction}

The relaxation to equilibrium of closed quantum systems and their temporal evolution are important areas of study, as seen in, for example~\cite{thermal1,thermal2,thermal3}, with a beautiful experiment by Kinoshita \etal~\cite{Kinoshita}  having shown that relaxation to equilibrium does not happen in a trapped one dimensional Bose gas with contact interactions. This was not unexpected for a one dimensional untrapped Bose gas with point interactions, which is known to be an integrable system, but it had been thought that practical features such as the harmonic trap and imperfectly point-like interactions would compromise the integrability and the system would relax. On the other hand, there are closed quantum systems which are known to relax to a generalised equilibrium, without any interactions with a thermal cloud or other reservoir~\cite{thermal2,zhang,lfs}.
To the best of our knowledge, there is as yet no established consensus on the mechanism by which this happens.

For the relaxation of closed quantum systems to a generalised equilibrium, one proposal is the eigenstate thermalisation hypothesis (ETH), in which every eigenstate of the Hamiltonian implicitly contains a thermal state~\cite{srednicki,thermal3}, with the initial coherence between the eigenstates being lost by dephasing during the dynamics. Srednicki, when introducing this hypothesis, claimed that a necessary condition was the validity of Berry's conjecture that the energy eigenfunctions behave like Gaussian random variables~\cite{Berry}, which is expected to hold for systems which exhibit classical chaos in at least a large majority of the classical phase space.

In this work we study the relaxation or lack thereof in bosonic ultra-cold atomic systems held in two~\cite{Joel}, three~\cite{Nemoto}, and four-well~\cite{Anglin,nosso} tunnel-coupled potentials. For these sytems, the only constants of the motion are the total number of atoms and the total energy, so that in the classical sense we would expect only the twin well model to be integrable. Following the general wisdom, we would therefore not expect the two well system to relax to equilibrium, whereas we have previously shown that the four-well system, at least in one particular configuration, will demonstrate this feature over medium times~\cite{nosso}.

We use stochastic integration in the truncated Wigner representation~\cite{Wminus1,Steel} to perform phase-space analyses of the quantum dynamics and calculate the expectation values of the well populations and other operator products which allow us to construct effective reduced single-particle density matrices. From these matrices we are able to calculate a pseudoentropy which we previously found useful for the four well system~\cite{nosso}. For the two-site model we are also able to calculate exact quantum results using a matrix equation for the number state coefficients, finding excellent agreement with the stochastic results over most of the time domain. We also calculate a type of Lyupanov exponent classically, using coupled Gross-Pitaevskii equations, and investigate the extent to which this allows us to predict which systems will relax. We find that all the systems we investigate will reach a generalised, but not necessarily static, equilibrium, but not for arbitrary initial quantum states. We find that the differences can be predicted qualitatively via the off-diagonal elements of the reduced density matrices at the beginning of the time evolution.

\section{Physical model, Hamiltonian and equations of motion}
\label{sec:model}

We consider three different physical arrangements, firstly with two wells of equal depth, secondly with three equal wells at the vertices of an equilateral triangle, and thirdly with four equal wells in a square arrangement. We will follow a procedure which is effectively equivalent to changing the Hamiltonians at $t=0$, in that all our initial conditions will have at least one well unoccupied.
We outline here the the standard procedure for developing an effective Hamiltonian for two wells~\cite{Joel}, where we consider separate wells with an independent condensate in each of the at the beginning of our investigations. The procedures for three or four wells are a simple extension of this~\cite{nosso}. The Hamiltonian for a condensate in an external trapping potential, $V_{ext}(\vec{r})$, may be written as
\begin{equation}
\hat{\cal{H}}=\int{ d \vec{r} \left[ \frac{\hbar^2}{2m} \nabla \hat{\psi}^{\dagger} \cdot \nabla \hat{\psi} +V_{ext}(\vec{r}) + U_0\hat{\psi}^{\dagger} \hat{\psi}^{\dagger} \hat{\psi} \hat{\psi}   \right] } , 
\label{eq:hamiltonian}
\end{equation}
where $\hat{\psi}$ is the field operator for the condensate, and the non-linear interaction parameter is $U_0=2\pi a \hbar^{2} / m$,  where $a$ is the s-wave scattering length describing two-body collisions within the condensate, and $m$ is the atomic mass. In the case where the external potential provides a two well confinement for the condensate, we may simplify the above Hamiltonian by making use of the two-mode approximation. At zero temperature all atoms in the system are condensed and if the ground state energies of the condensate in either well are sufficiently separated from the energies of the condensate in all other excited single particle states, transitions to or from the modes of interest and these higher lying states can be neglected. We may then expand the field operator as
\begin{equation}
\hat{\psi}(\vec{r})\approx  \left(\phi_{1}(\vec{r}) \hat{a}_{1} + \phi_{2}(\vec{r})\hat{a}_{2}\right) ,
\label{eq:2modeoperators}
\end{equation}
where the $\hat{a}_{i}$  are bosonic annihilation operators in each of the wells, and the $\phi_{i}$ are the ground state spatial wave functions of the condensate in each of the wells.

Using this in Eq.~(\ref{eq:hamiltonian}), we find an effective Hamiltonian
\begin{eqnarray}
\hat{\cal{H}}_{eff} &=& E_{1} \hat{a}_{1}^\dagger \hat{a}_{1}+ E_{2} \hat{a}_{2}^\dagger \hat{a}_{2}\nonumber\\
& & + \hbar\chi \left(\hat{a}_{1}^\dagger \hat{a}_{1}^\dagger \hat{a}_{1} \hat{a}_{1}+ \hat{a}_{2}^\dagger \hat{a}_{2}^\dagger \hat{a}_{2} \hat{a}_{2}\right)\nonumber\\
& &
-\hbar J \left(\hat{a}_{1}^\dagger \hat{a}_{2}+\hat{a}_{2}^\dagger \hat{a}_{1} \right),
\label{eq:effhamiltonian}
\end{eqnarray}
where we have neglected the spatial overlap of the different well densities. The single well bound state energies, $E_{i}$, are
\begin{equation}
E_{i} = \int{d\vec{r}~ \phi_{i}^{\ast}(\vec{r}) \left(\frac{-\hbar^2}{2m}\nabla^2 + V_{ext}(\vec{r}) \right) \phi_{i}(\vec{r}) }.
\label{eq:wellenergies}
\end{equation}
$J$, the tunnel coupling, is
\begin{equation}
J = \frac{-1}{\hbar}\int{d\vec{r}~ \phi_{1}^{\ast}(\vec{r}) \left(\frac{-\hbar^2}{2m}\nabla^2 + V_{ext}(\vec{r}) \right) \phi_{2}(\vec{r}) } ,
\label{eq:Jdeff}
\end{equation}
and effective non-linear interaction term is
\begin{equation}
\chi = \frac{U_0}{\hbar} \int{d\vec{r}~ \vert \phi_{i}(\vec{r}) \vert^4 } .
\label{eq:interactioneqn}
\end{equation}
We set the single well bound state energies equal because we will consider only symmetric potentials where we can set $E_{1}=E_{2}=0$. 

\begin{figure}
\begin{center}
\includegraphics[width=0.45\columnwidth]{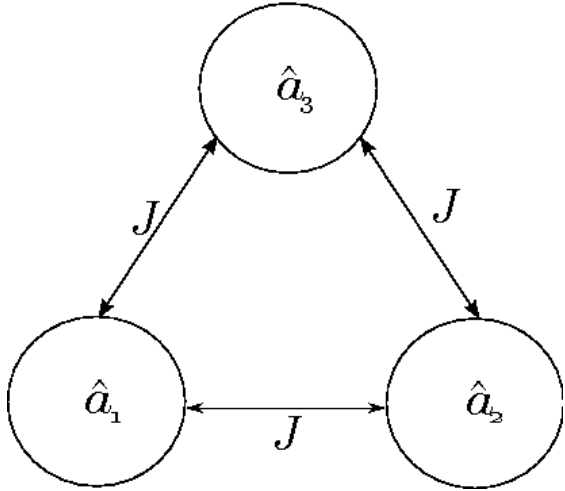}
\end{center}
\caption{Schematic of our three-mode Bose-Hubbard system. The $\hat{a}_{i}$ are the bosonic annihilation operators for each mode, while $J$ represents the coupling rate between the modes.  In this article, we always set $J=1$, which sets the units of time.}
\label{fig:triangle}
\end{figure}

\begin{figure}
\begin{center}
\includegraphics[width=0.45\columnwidth]{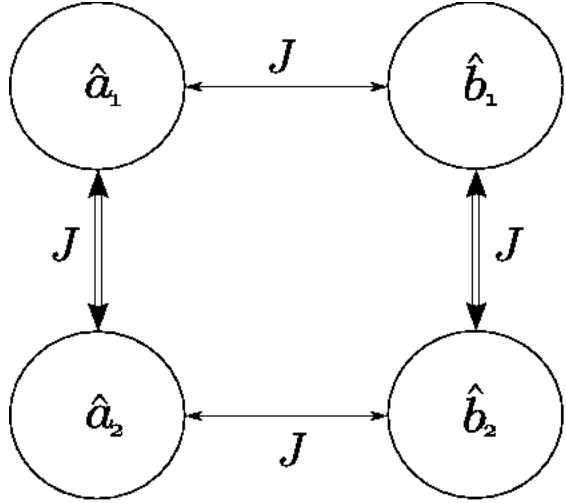}
\end{center}
\caption{Schematic of our four-mode Bose-Hubbard system. The $\hat{a}_{i}$ and $\hat{b}_{i}$ are the bosonic annihilation operators for each mode, while $J$ represents the coupling rate between the modes.}
\label{fig:fourmode}
\end{figure}

This process simplifies the analyses by approximating the atoms in each separate well as being in a single mode, meaning that our equations are much easier to solve than would otherwise be the case.
Generalising to the three and four site models, we will use $\hat{a}_{i}\:\:(i=1,2,3)$ as bosonic annihilation operators in each of the two or three wells, and $\hat{a}_{i}$ and $\hat{b}_{i}$, with $i=1,2$, in each side of the four-well system. We parametrise time by setting $J=1$, so that dimensionless time as displayed in the results is labelled as $Jt$. A schematic of the three well configuration is shown in Fig.~\ref{fig:triangle}, with the four well system being treated as shown in Fig.~\ref{fig:fourmode}.

We may now also write the effective Hamiltonians for the three and four site systems, generalising from that for twin wells.
For the three-well system we find the effective Hamiltonian,
\begin{eqnarray}
\hat{\cal{H}}_{eff}^{(3)} &=& \displaystyle\sum\limits_{i=1}^{3} \left(\hbar\chi \hat{a}_{i}^\dagger 
\hat{a}_{i}^\dagger \hat{a}_{i} \hat{a}_{i}\right)\nonumber\\
& &
-\hbar J\left(\hat{a}_{1}^\dagger \hat{a}_{2}+\hat{a}_{2}^\dagger \hat{a}_{1}+
\hat{a}_{1}^\dagger \hat{a}_{3}+\hat{a}_{3}^\dagger \hat{a}_{1}+\hat{a}_{2}^\dagger \hat{a}_{3}+\hat{a}_{3}^\dagger \hat{a}_{2}\right).
\label{eq:effham3}
\end{eqnarray}
Finally, for the the four-well system we have,
\begin{eqnarray}
\hat{\cal{H}}_{eff}^{(4)} &=& \displaystyle\sum\limits_{i=1}^{2} \left( \hbar\chi \hat{a}_{i}^\dagger \hat{a}_{i}^\dagger \hat{a}_{i} \hat{a}_{i}+\hbar\chi \hat{b}_{i}^\dagger \hat{b}_{i}^\dagger \hat{b}_{i} \hat{b}_{i}\right)\nonumber\\
& &
-\hbar J \left(\hat{a}_{1}^\dagger \hat{a}_{2}+\hat{a}_{2}^\dagger \hat{a}_{1}+\hat{b}_{1}^\dagger \hat{b}_{2}+\hat{b}_{2}^\dagger \hat{b}_{1}+\hat{a}_{1}^\dagger \hat{b}_{1}+\hat{b}_{1}^\dagger \hat{a}_{1}+\hat{a}_{2}^\dagger \hat{b}_{2}+\hat{b}_{2}^\dagger \hat{a}_{2}\right).
\label{eq:effham4}
\end{eqnarray}

In order to calculate the quantum dynamics of these three systems, we will make use of the truncated Wigner representation~\cite{Wminus1,Steel}, which gives results indistinguishable from those of the full quantum matrix equations for the number state coefficients~\cite{DanChris} up until at least the relaxation time in the case of two and three wells, and is much easier to use for four-wells, where the full matrix equations become very difficult to use. Following the standard methods~\cite{QNoise}, we find generalised Fokker-Planck equations for the Wigner pseudoprobability functions. As these contain third order derivatives, they cannot be mapped onto stochastic differential equations. Although methods have been developed to allow a mapping onto stochastic difference equations~\cite{nossoEPL}, the numerical integration of these equations is even more unstable than that of the positive-P representation equations~\cite{Steel,P+}, so we will not follow this route here.
Instead, we truncate the third order terms in the generalised Fokker-Planck equations and make a mapping to coupled equations for the Wigner variables corresponding to the operators found in the Hamiltonians. We note here that classical averages of the Wigner variables correspond to
symmetrically ordered operator expectation values, so that the
necessary reordering must be undertaken before we arrive at solutions
for physical quantities, for which normal ordering is more appropriate.

Having done this, we find the sets of coupled equations given immediately below. For the two-well system, the truncated Wigner equations of motion are,
\begin{eqnarray}
\frac{d\alpha_{1}}{dt} &=& -2i\chi|\alpha_{1}|^{2}\alpha_{1} +iJ\alpha_{2}\nonumber\\
\frac {d \alpha_{2}}{dt} &=&
-2i\chi|\alpha_{2}|^{2}\alpha_{2} +iJ\alpha_{1},
\label{eq:Wminus2}
\end{eqnarray}
for the three-well system they are
\begin{eqnarray}
\frac {d \alpha_{1}}{dt} &=&
-2i\chi|\alpha_{1}|^{2}\alpha_{1} +iJ\left(\alpha_{2}+\alpha_{3}\right) \nonumber\\
\frac {d \alpha_{2}}{dt} &=&
-2i\chi|\alpha_{2}|^{2}\alpha_{2} +iJ\left(\alpha_{1} + \alpha_{3}\right)\nonumber\\
\frac {d \alpha_{3}}{dt} &=&
-2i\chi|\alpha_{3}|^{2}\alpha_{3} +iJ\left(\alpha_{1} + \alpha_{2}\right),
\label{eq:Wminus3}
\end{eqnarray}
and for the four-wells they are
\begin{eqnarray}
\frac {d \alpha_{1}}{dt} &=&
-2i\chi|\alpha_{1}|^{2}\alpha_{1} +iJ\left(\alpha_{2} + \beta_{1}\right)\nonumber\\
\frac {d \alpha_{2}}{dt} &=&
-2i\chi|\alpha_{2}|^{2}\alpha_{2} +iJ
\left(\alpha_{1} + \beta_{2}\right)\nonumber\\
\frac {d \beta_{1}}{dt} &=&
-2i\chi|\beta_{1}|^{2}\beta_{1} +iJ\left(\beta_{2} + \alpha_{1}\right)\nonumber\\
\frac {d \beta_{2}}{dt} &=&
-2i\chi|\beta_{2}|^{2}\beta_{2} +iJ\left(\beta_{1} +  \alpha_{2}\right).
\label{eq:Wminus4}
\end{eqnarray}

Although the three sets of equations above might look
classical, the Wigner variables themselves are drawn from appropriate
distributions for the desired initial quantum states, with the
stochasticity coming from the initial conditions for the chosen quantum states. The truncated Wigner equations are solved numerically by taking averages over a large number of stochastic trajectories, with initial conditions sampled probabilistically~\cite{nosso,AxMuzza}.

\section{Classical integrability and stability}
\label{sec:Lyupanov}

Before we investigate the quantum dynamics further we will examine the classical systems, which are described by coupled Gross-Pitaevskii equations which have the same form as above, but have deterministic initial conditions. Berry's conjecture~\cite{Berry} states that a quantum system can thermalise if the related classical system is unstable or chaotic in at least a large majority of the classical phase space.   
To examine the applicability of this conjecture, we numerically calculate approximate Lyupanov exponents for the three and four mode systems~\cite{Eckmann}, which we define for the dynamics of one of the coupled wells as
\begin{equation}
L_{j} = \lim_{\tau\rightarrow\infty}\frac{1}{\tau}\frac{\ln\left(\delta N_{j}(\tau)\right)}{\delta N_{j}(0)},
\label{eq:Lyapdef}
\end{equation}
where
\begin{equation}
\delta N_{j}(\tau)=|N_{j}^{(1)}(\tau)-N_{j}^{(2)}(\tau)|,\:\:\:j=1,2,3,4,
\label{eq:deltadef}
\end{equation}
where $N_{j}^{(2)}$ is an initial condition slightly perturbed from $N_{j}^{(1)}$. Note that we have used numbers here rather than complex amplitudes, as it is the numbers which are directly observable. In practice, we obviously cannot integrate the equations for infinite time, so we integrate the coupled GPE type equations over a reasonably long time and look at the development of $\delta N_{j}(t)$ and hence $L_{j}(t)$. We also make the caveat here that because the system is both bounded and periodic, these effective Lyupanov exponents are not as definitive as in a non-periodic system.

In these systems as we define them, the only constants of the motion are the total energy and the total number of atoms. The obvious classical degrees of freedom are the number of atoms in each well, which, given the constraint on total number, would seem to be one less than the number of wells. This approach suggests that the three-well system has the same number of classical degrees of freedom as it does constants of the motion, and therefore should be classically integrable. The four-well system, as we have previously shown, is not stable, at least when two different tunneling rates are present~\cite{nosso}.

\begin{figure}
\begin{center}
\subfigure{\includegraphics[width=0.45\columnwidth]{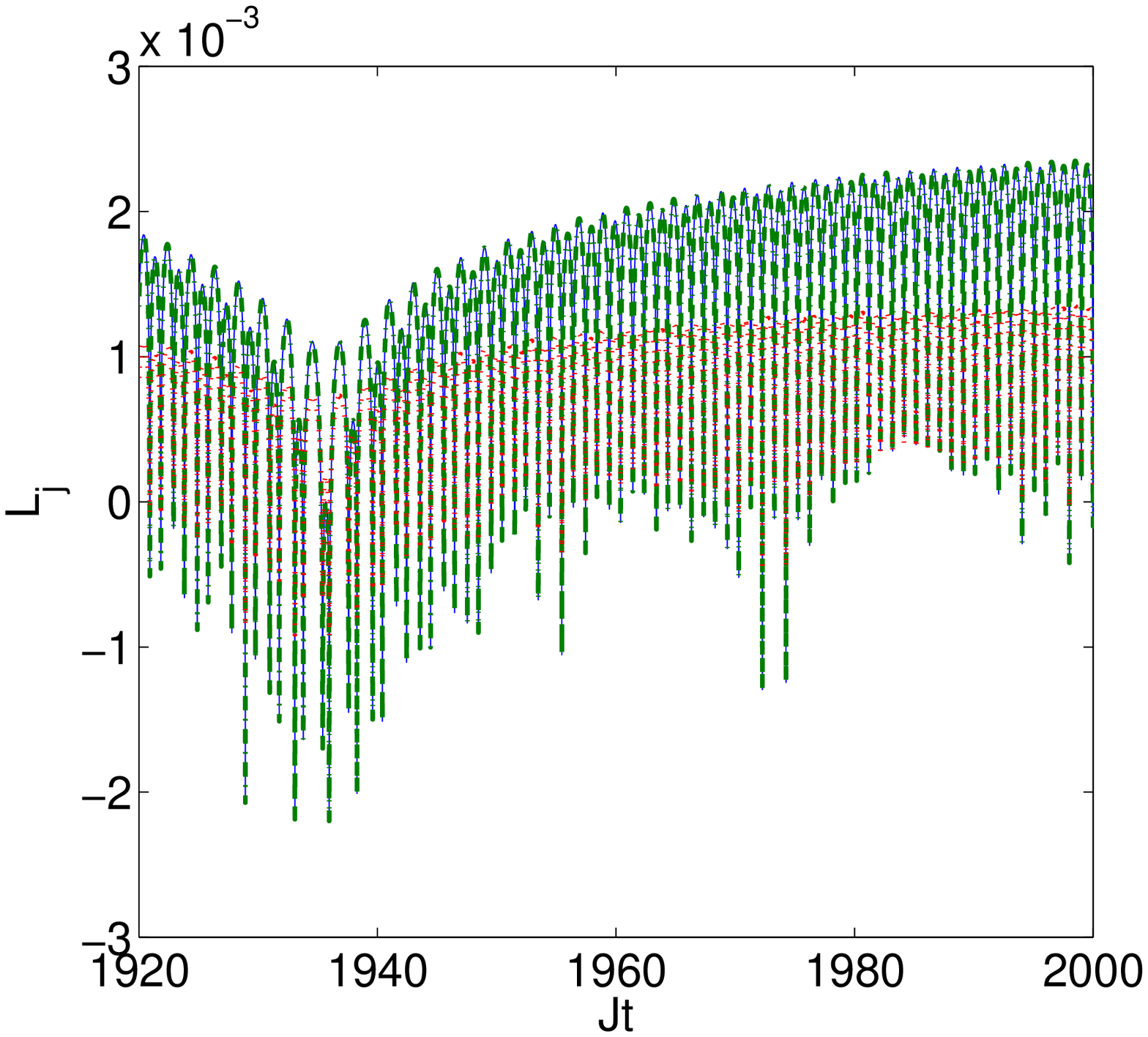}
\label{fig:Ly3}}
\hspace{8pt}
\subfigure{\includegraphics[width=0.45\columnwidth]{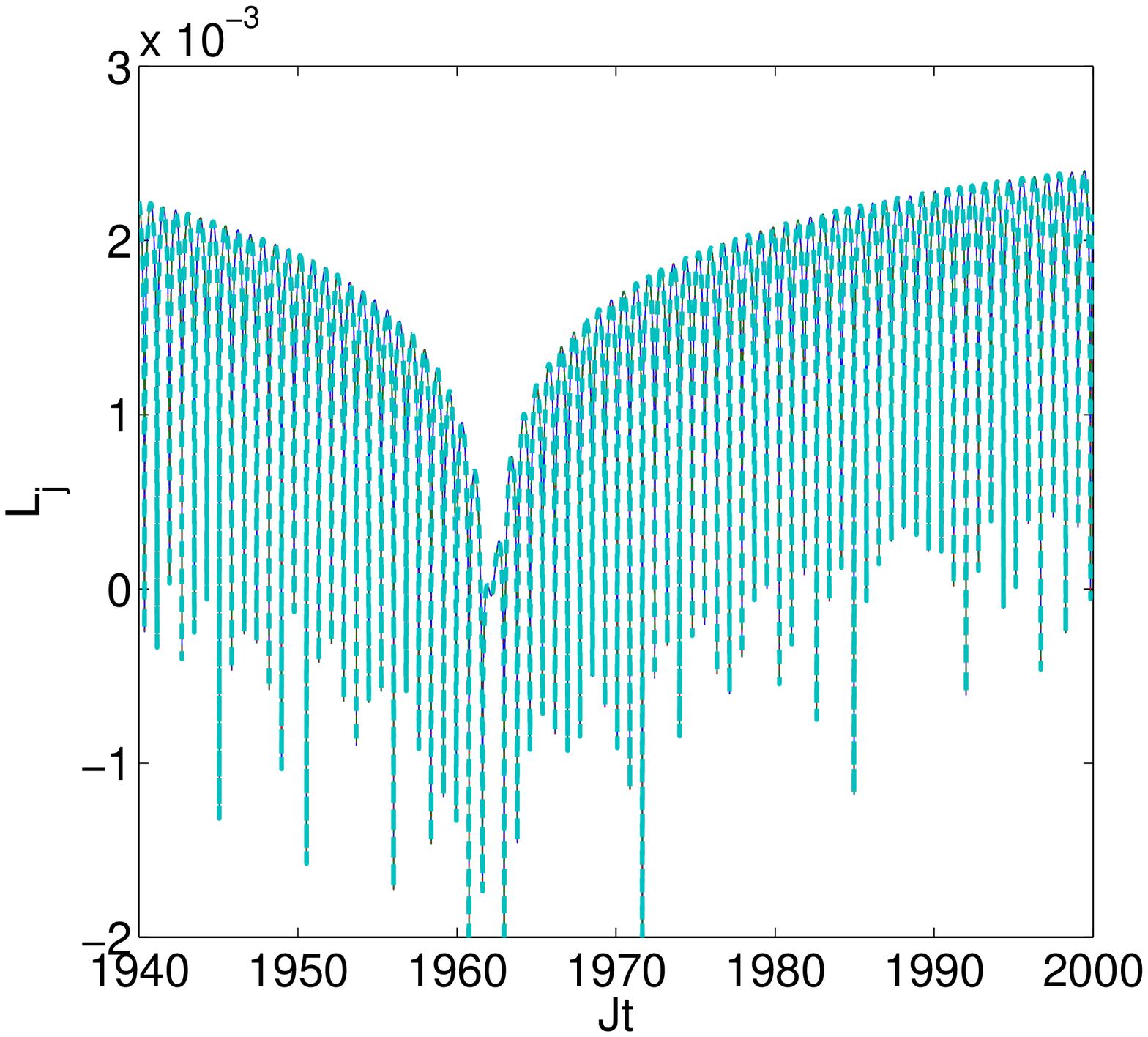}
\label{fig:Ly4}}
\end{center}
\caption{(colour online) The time development of the Lyupanov exponents for the three- and four-well systems. For \subref{fig:Ly3} the reference initial condition was with all $300$  atoms equally distributed between wells one and two, with none in well three, and the perturbed initial condition had $149$ in wells one and two, with $2$ atoms in well three. $\chi N_{tot} = 0.5J$, with $J=1$. The solid line is $L_{1}$, the dash-dotted is $L_{2}$ (dynamics indistinguishable from $L_{1}$) and the dotted line is $L_{3}$. For \subref{fig:Ly4}, the reference initial condition was with $200$  atoms in each of wells $a_{1}$ and $b_{2}$, with the other two unoccupied.  The perturbed initial condition had $199$ in wells $a_{1}$ and $b_{2}$, with $1$ atom in each of the other two. $\chi N_{tot} = 0.5J$, with $J=1$. The four exponents are indistinguishable. All quantities plotted in this and subsequent plots are dimensionless.}
\label{fig:Lyupanov}
\end{figure}

In Fig.~\ref{fig:Lyupanov}\subref{fig:Ly3} we show the exponents for the three-well system, beginning in a situation far from the expected equilibrium, which would have one third of the atoms in each well. The classical solutions are again regular and periodic, with full oscillations between $0$ and $300$ atoms in well three, with the other two oscillating between $150$ and $0$. The perturbed classical solutions, while still regular and periodic, never see more than $149$ or less than $1$ atom in wells one and two, while well three oscillates between $2$ and $302$. Fig.~\ref{fig:Lyupanov}\subref{fig:Ly4} shows that the four-well system also behaves in a similar manner.

\section{Relaxation}
\label{sec:thermal}

In a state of relaxed equilibrium, we expect that the entropy of the system will be maximised. In a quantum system, the entropy is normally defined using the density matrix for the system. While this works very well for small systems, such as those that are of interest in discrete variable quantum information applications, it becomes difficult to calculate the full density matrix for a many-body interacting quantum system, which will have a much larger Hilbert space. As in our previous work~\cite{nosso}, we can define something which behaves as a reduced single-particle density matrix, which then allows us to calculate an effective entropy. Importantly for possible experimental investigations, all the quantities needed are in principle measurable using techniques developed by Ferris \etal~\cite{Andyhomo}. For the four-well system we define, 
\begin{eqnarray}
\rho_{4} = \frac{1}{\langle \hat{N}_{tot}\rangle}\left(\begin{array}{cccc}
\langle \hat{a}_{1}^{\dag}\hat{a}_{1}\rangle & \langle \hat{a}_{1}^{\dag}\hat{a}_{2}\rangle & \langle \hat{a}_{1}^{\dag}\hat{b}_{1}\rangle & \langle \hat{a}_{1}^{\dag}\hat{b}_{2}\rangle
\\
\langle \hat{a}_{2}^{\dag}\hat{a}_{1}\rangle & \langle \hat{a}_{2}^{\dag}\hat{a}_{2}\rangle & \langle \hat{a}_{2}^{\dag}\hat{b}_{1}\rangle & \langle \hat{a}_{2}^{\dag}\hat{b}_{2}\rangle
\\
\langle \hat{b}_{1}^{\dag}\hat{a}_{1}\rangle & \langle \hat{b}_{1}^{\dag}\hat{a}_{2}\rangle & \langle \hat{b}_{1}^{\dag}\hat{b}_{1}\rangle & \langle \hat{b}_{1}^{\dag}\hat{b}_{2}\rangle 
\\
\langle \hat{b}_{2}^{\dag}\hat{a}_{1}\rangle & \langle \hat{b}_{2}^{\dag}\hat{a}_{2}\rangle & \langle \hat{b}_{2}^{\dag}\hat{b}_{1}\rangle & \langle \hat{b}_{2}^{\dag}\hat{b}_{2}\rangle
\end{array}\right),
\label{eq:nossomat}
\end{eqnarray}
with the matrices for two and three wells being obvious truncations of the above. It is then an easy matter to calculate a single-particle pseudoentropy from this matrix,
\begin{equation}
\xi_{j}(t) = -\mathrm{Tr}\left[\rho_{j}(t)\ln \rho_{j}(t)\right],
\label{eq:entropy}
\end{equation}
which will have a maximum value of $\ln j$ when the atoms are equally distributed throughout the $j$ wells, which is statistically the most probable situation.

\begin{figure}
\begin{center}
\subfigure{\includegraphics[width=0.45\columnwidth]{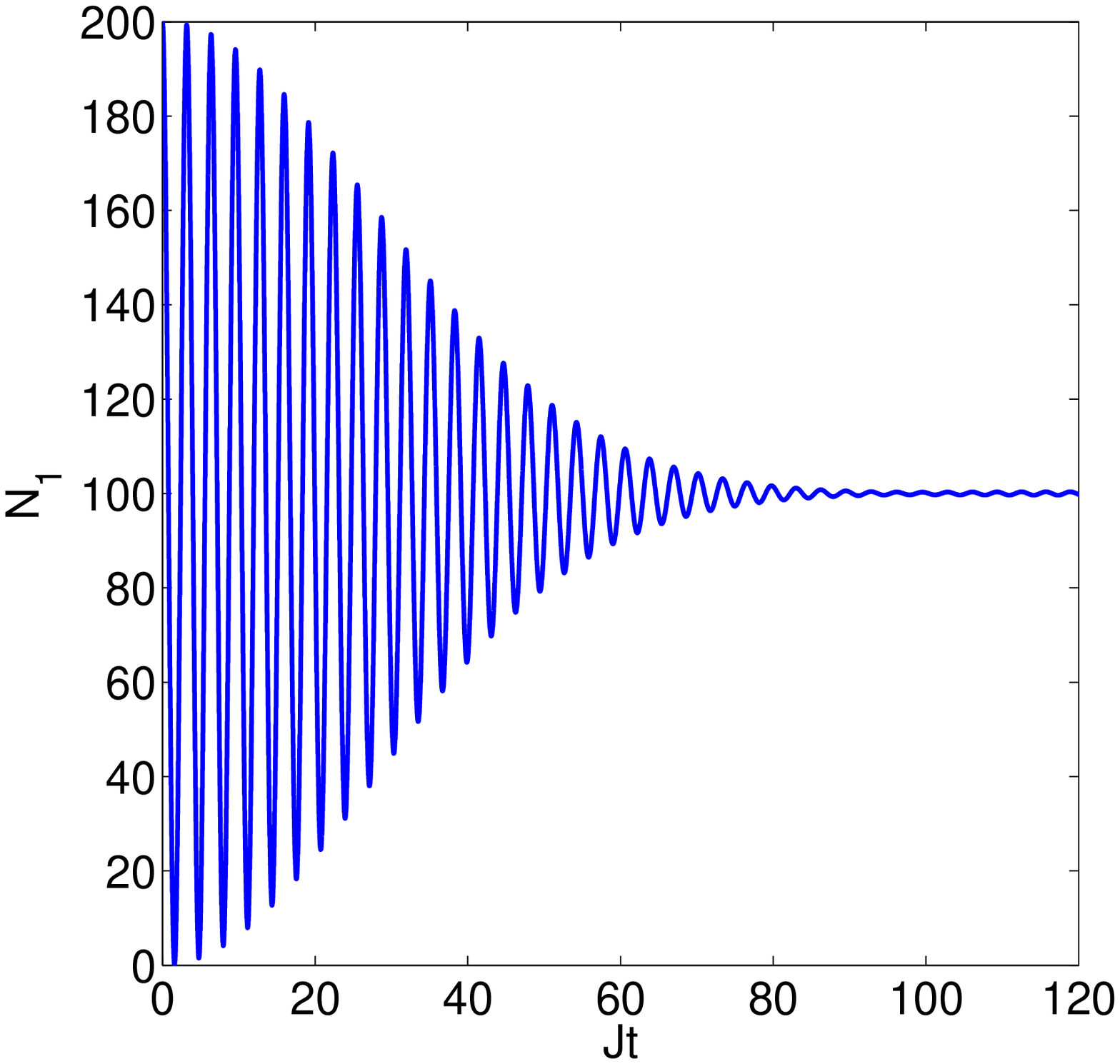}
\label{fig:number2}}
\hspace{8pt}
\subfigure{\includegraphics[width=0.45\columnwidth]{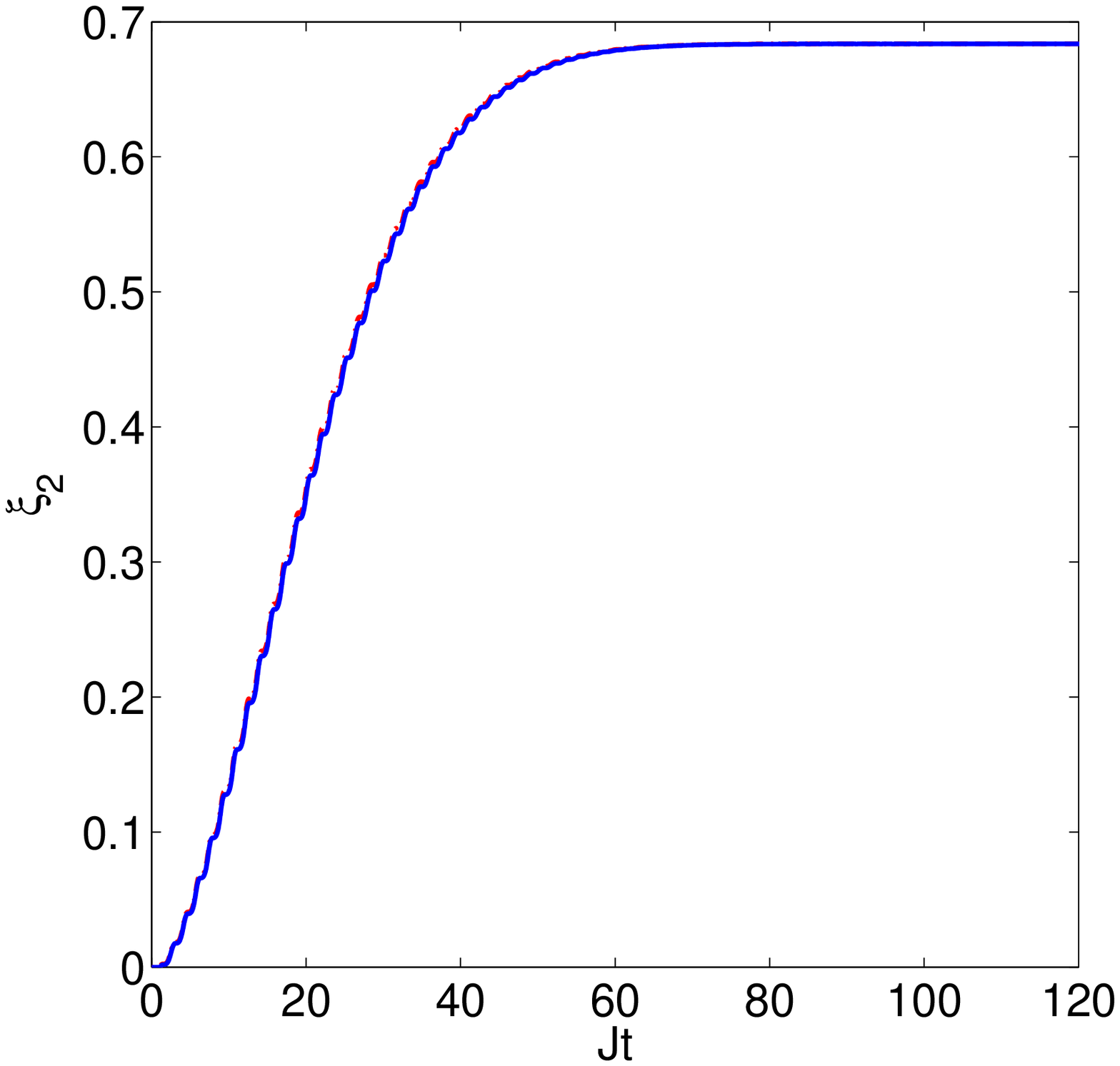}
\label{fig:entropy2}}
\end{center}
\caption{(colour online) The mean-field population dynamics and pseudoentropy for the two-well system, with $N_{1}(0)=200$, $N_{2}(0)=0$, $\chi N_{tot}=0.5J$, and $J=1$ , calculated using the truncated Wigner representation. The solid (blue) lines are the average over $1.57\times 10^{5}$ trajectories for an initial Fock state, while the dash-dotted lines (red) are averaged over $2.33\times 10^{5}$ trajectories for an initial coherent state.
\subref{fig:number2} gives the expectation values of the population in well one and \subref{fig:entropy2} gives the calculated pseudentropies.}
\label{fig:numberentropy2}
\end{figure}

\begin{figure}
\begin{center}
\subfigure{\includegraphics[width=0.45\columnwidth]{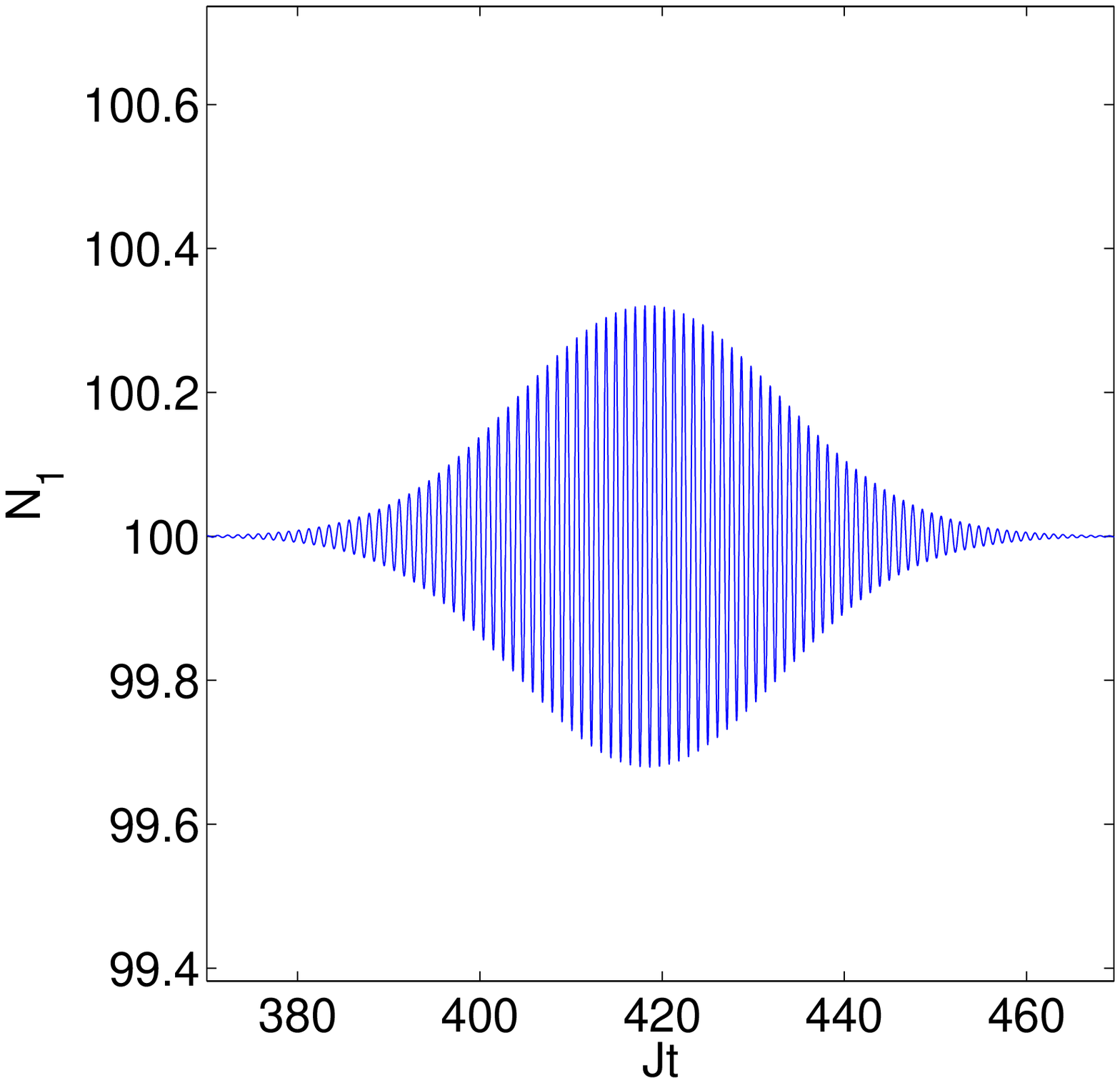}
\label{fig:revive2}}
\hspace{8pt}
\subfigure{\includegraphics[width=0.45\columnwidth]{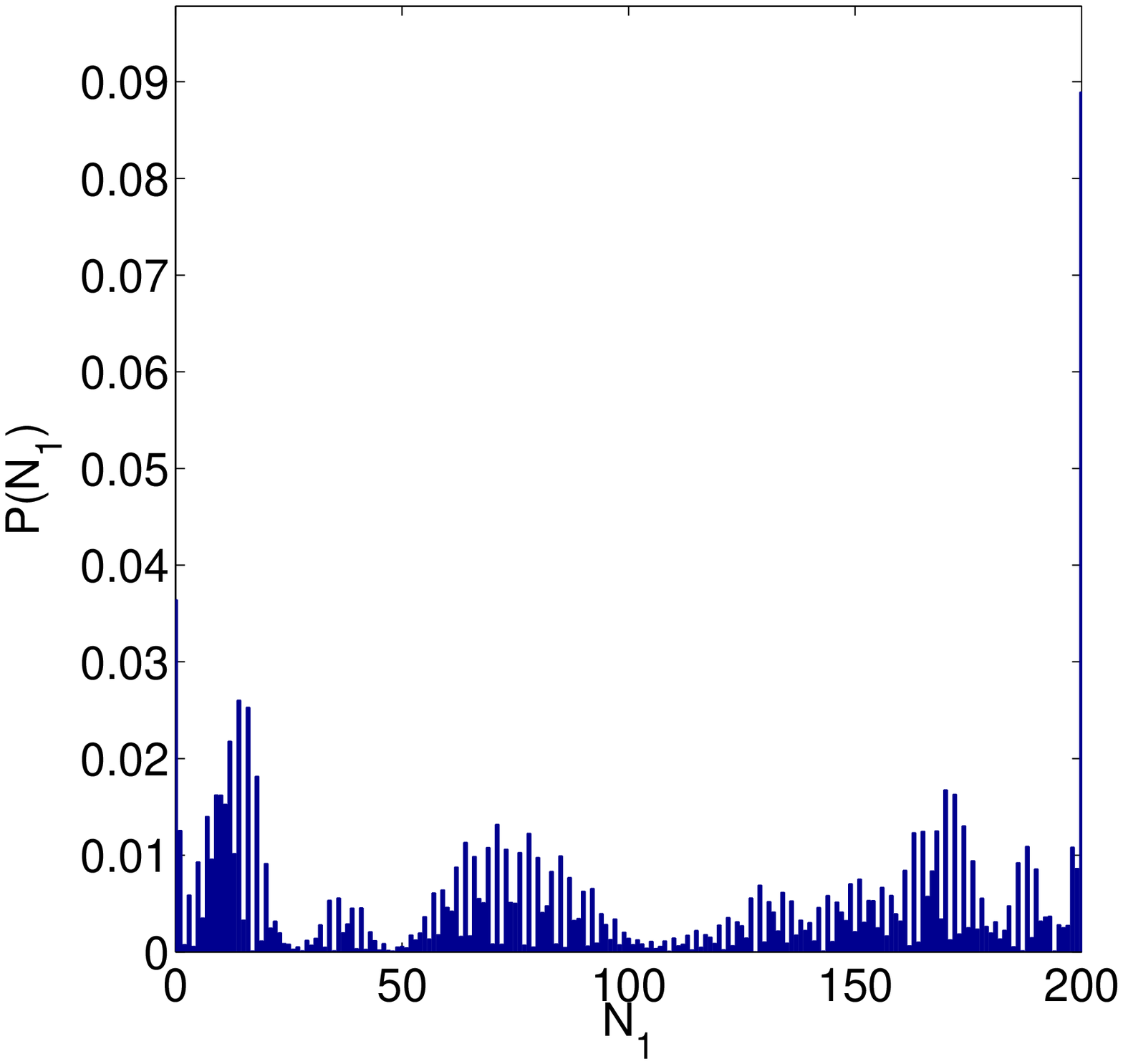}
\label{fig:prob2}}
\end{center}
\caption{(colour online) The first mean-field partial revival and the probablities for each number in well one at $Jt=900$, calculated using the exact method for the same parameters and initial conditions as in Fig.~\ref{fig:numberentropy2}, but with initial Fock states.}
\label{fig:bijordan2}
\end{figure}

In Fig.~\ref{fig:numberentropy2} we show the results for the two-well system, beginning with all $200$ atoms initially in one of the wells.  From Fig.~\ref{fig:numberentropy2}\subref{fig:number2} we see that the atoms equalise between the two wells, with the initial quantum states having almost no discernible effect on the process. This is also clear when we look at Fig.~\ref{fig:numberentropy2}\subref{fig:entropy2}, where the pseudoentropy rises to a value very close to $\ln 2\approx 0.6931$, which is the maximum single-particle value expected of this system in thermal equilibrium. We therefore see that, once quantum effects are included, this system appears to relax to something close to its equilibrium state at zero temperature without any contact with a reservoir, despite the fact that it is classically integrable.  
We have also integrated the exact quantum equations for the number state coefficients using a matrix method for initial Fock states~\cite{DanChris,Tania} and see a partial revival of the oscillations centred around $Jt=420$, as shown in Fig.~\ref{fig:bijordan2}\subref{fig:revive2}. We used this method out to $Jt=900$, finding one more very partial revival at around $Jt=800$. In Fig.~\ref{fig:bijordan2}\subref{fig:prob2}, we show the $P(N_{1})$ at $Jt=900$ for well one of this system. The true equilibrium ground state for repulsive interatomic interactions would be a binomial distribution between the two wells, with the distribution in each well closely approximating a Gaussian centred on $N_{j}=100$. We see that this is not the case here, with the highest probably actually being for all the atoms in well one. This shows that, despite the pseudoentropy being close to its maximum and the mean populations being equal at this time, the system is not in a true equilibrium state. We will return to this issue below.    

\begin{figure}
\begin{center}
\subfigure{\includegraphics[width=0.45\columnwidth]{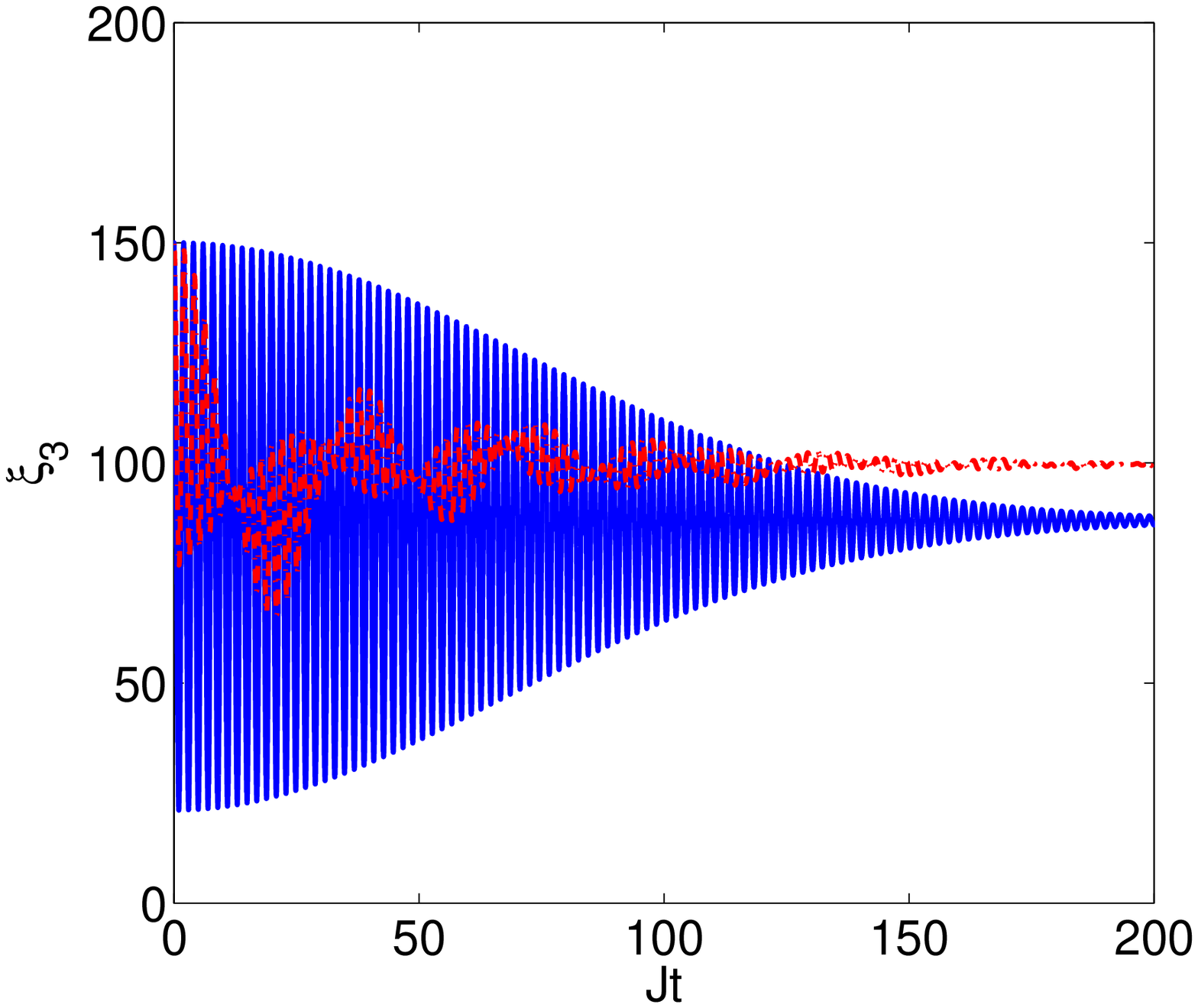}
\label{fig:number3}}
\hspace{8pt}
\subfigure{\includegraphics[width=0.45\columnwidth]{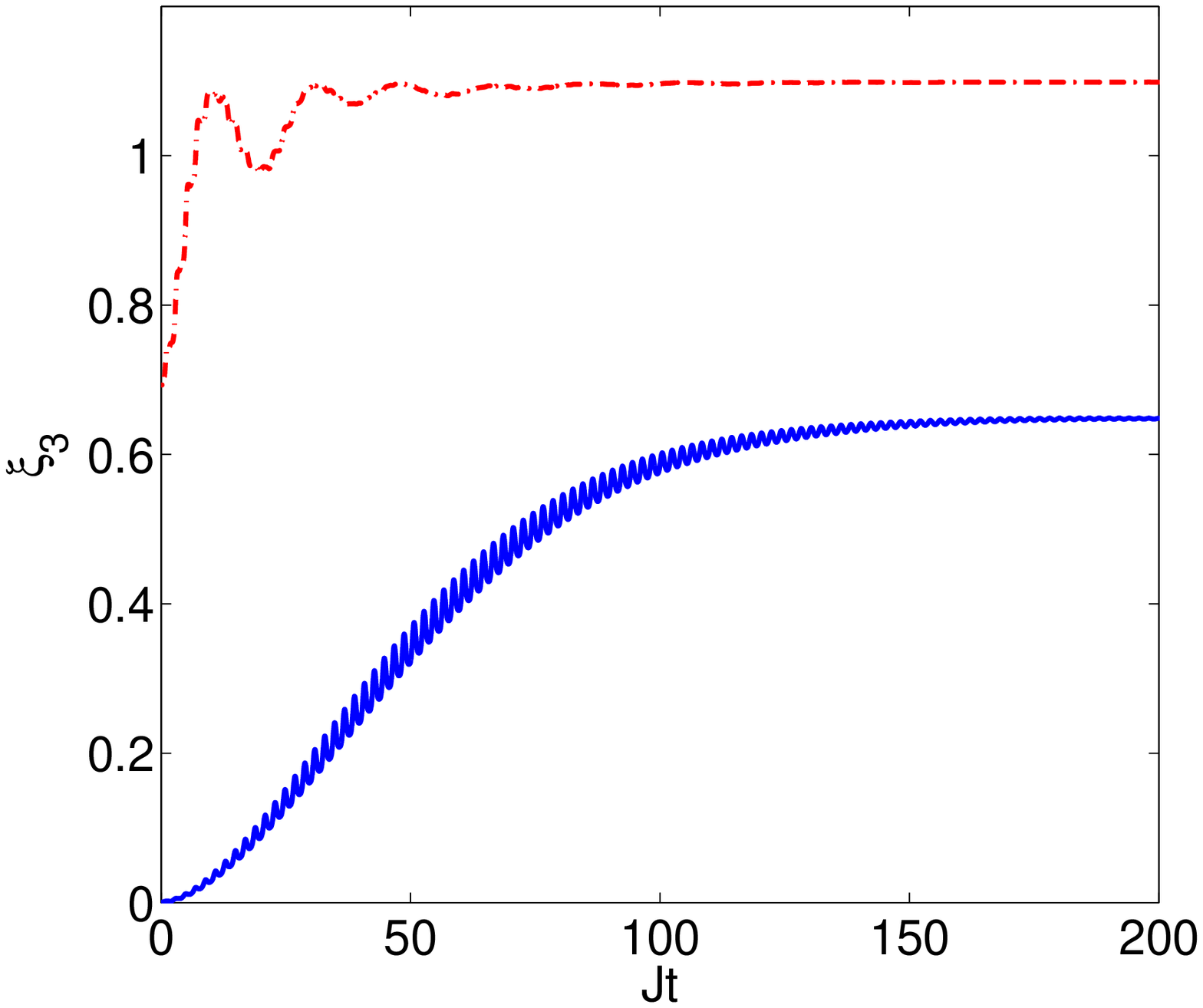}
\label{fig:entropy3}}
\end{center}
\caption{(colour online) The mean-field population dynamics of well $1$ and the pseudoentropy for the same parameters and initial conditions as the reference state in Fig.~\ref{fig:Lyupanov}\subref{fig:Ly3} in the three-well model. The solid (blue) lines are the average over $1.04\times 10^{9}$ trajectories for initial coherent states, while the dash-dotted lines (red) are averaged over $8\times 10^{4}$ trajectories for initial Fock states. The total atom number was $300$, with half of these initially in well $1$ and half in well $2$.
\subref{fig:number3} gives the expectation values of the population in well one and \subref{fig:entropy3} gives the calculated pseudoentropies.}
\label{fig:numberentropy3}
\end{figure}

When we look at the triple well system, we find differences which depend entirely on the quantum statistics, in a similar manner to previous investigations of quantum superchemistry~\cite{RCsuper,Wigsuper}. As shown in Fig.~\ref{fig:numberentropy3}, when we start with a total of $300$ atoms, half in one well and half in another, with the third well unoccupied, the subsequent behaviours for initial coherent and Fock states are qualitatively different. The initial Fock states equilibriate to an equal number in each well and the pseudoentropy climbs to very close to $\ln 3 \approx 1.0986$, but initial coherent states evolve to a different final distribution. We find approximately $87$ atoms in each the wells which were originally occupied, while the initially unoccupied well holds a final number of $126$. The final value of the pseudoentropy is well below the possible maximum, as the state of the three wells at the end of our integration time is by no means the statistically most probable distribution. A clue to the explanation of this behaviour can be found in the reduced single-particle density matrices at the final time of $Jt=200$. For the initial Fock states we find,
\begin{eqnarray}
\rho_{3} = \left(\begin{array}{ccc}
0.3315  & 0.0031 - 0.0017i & -0.0023 + 0.0001i \\
0.0031 + 0.0017i &  0.3297 & -0.0019 - 0.0019i \\
-0.0023 - 0.0001i &  -0.0019 + 0.0019i &  0.3388             
\end{array}\right),
\label{eq:Fockrho}
\end{eqnarray} 
while for initial coherent states the final result is
\begin{eqnarray}
\rho_{3} = \left(\begin{array}{ccc}
0.2895    &    0.2880 + 0.0000i &  0.1039 + 0.0004i \\
0.2880 - 0.0000i  &  0.2897  &  0.1039 + 0.0003i \\
0.1039 - 0.0004i  & 0.1039 - 0.0003i  & 0.4209 
\end{array}\right),
\label{eq:cohrho}
\end{eqnarray} 
from which we see that the off-diagonal elements are much larger for the initial coherent states. As these elements represent coherences between the modes, they contain information and therefore, the larger they are, the further the system is from the maximum entropy equilibrium state. 

\begin{figure}
\begin{center}
\subfigure{\includegraphics[width=0.45\columnwidth]{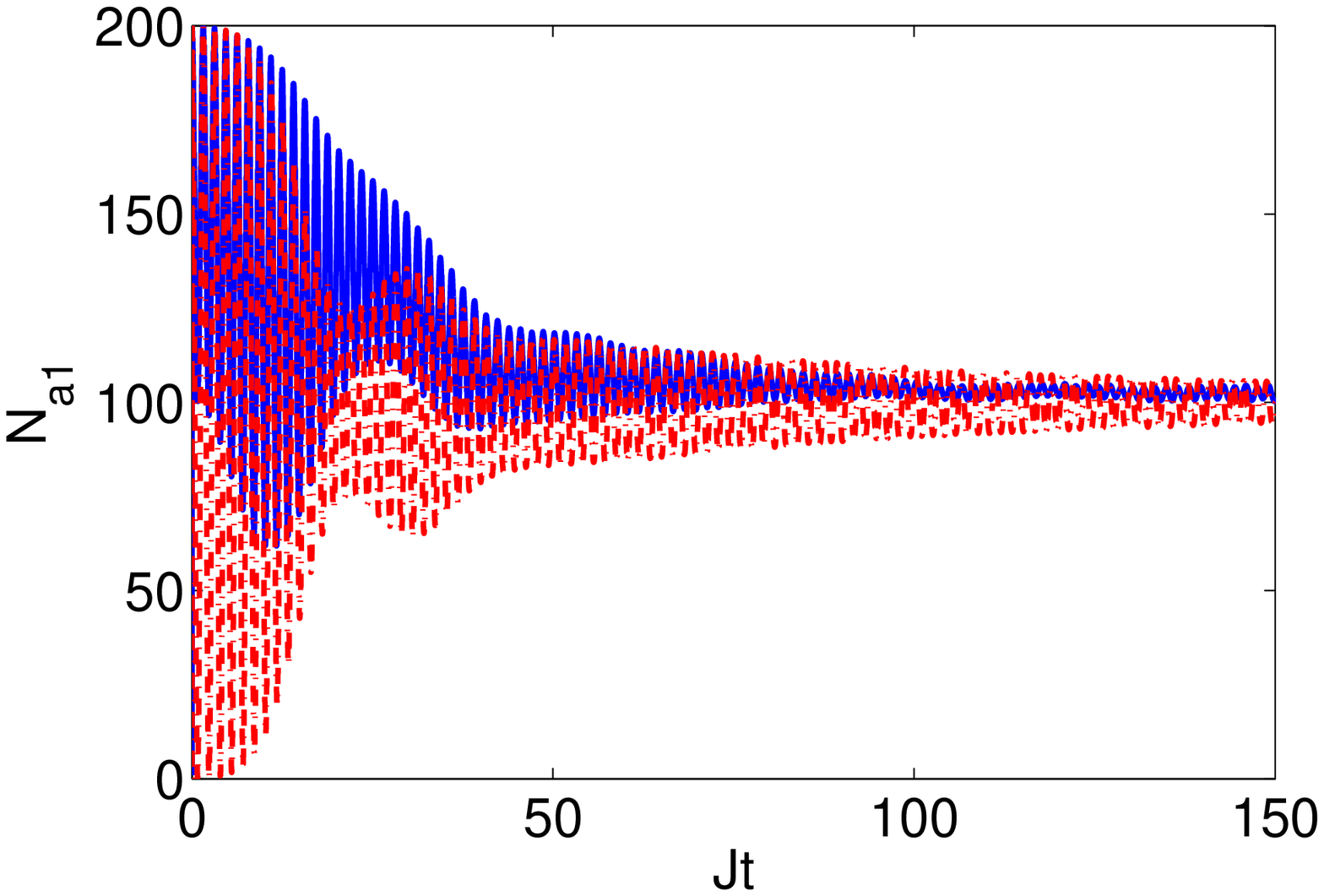}
\label{fig:number4}}
\hspace{8pt}
\subfigure{\includegraphics[width=0.45\columnwidth]{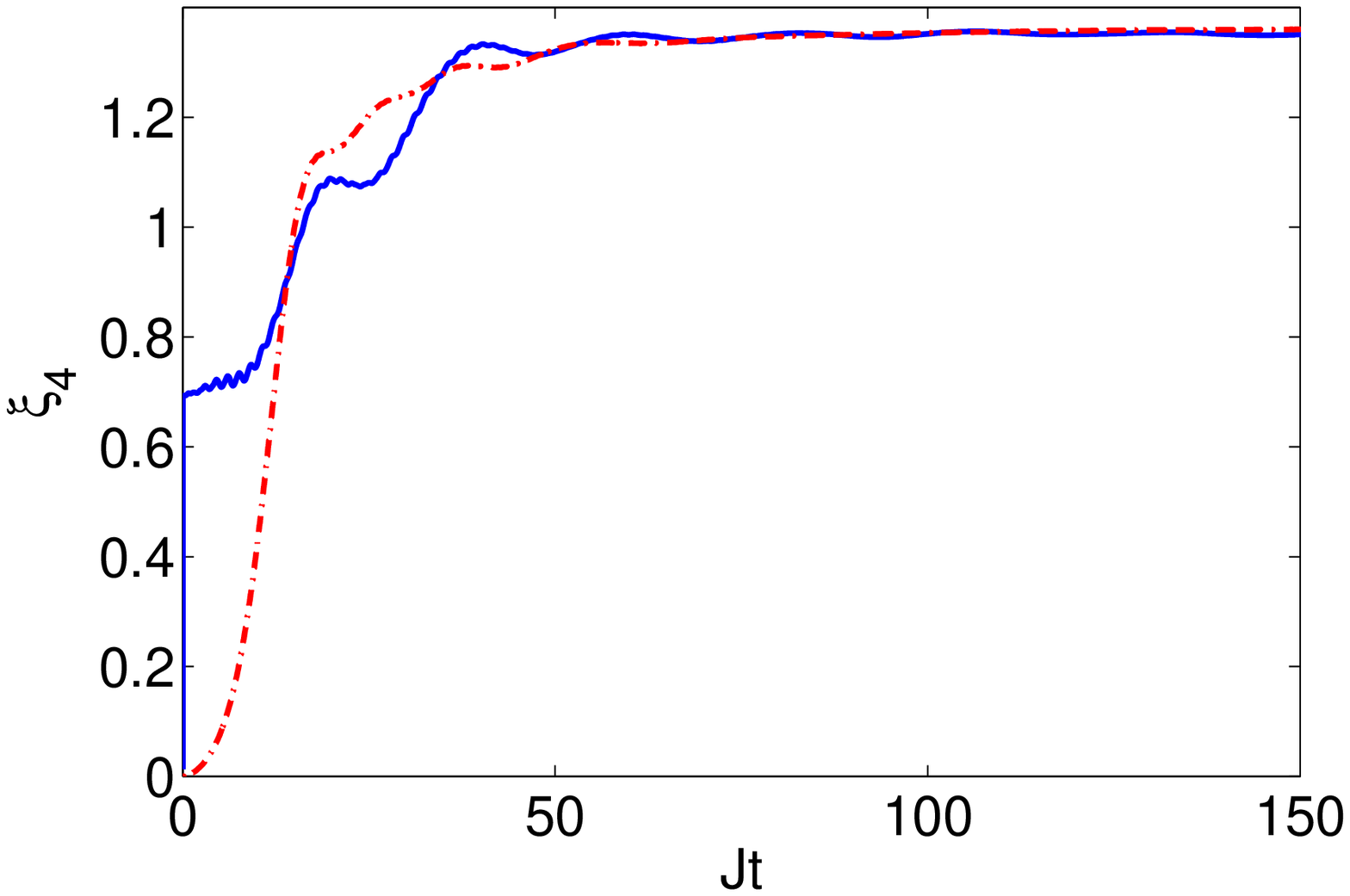}
\label{fig:entropy4}}
\end{center}
\caption{(colour online) The population dynamics of well $a_{1}$of the four-well system and the pseudoentropy for the same parameters and initial conditions as the reference state of Fig.~\ref{fig:Lyupanov}\subref{fig:Ly4}. The solid (blue) lines are the average over $1.36\times 10^{5}$ trajectories for initial Fock states, while the dash-dotted lines (red) are averaged over $4.05\times 10^{5}$ trajectories for initial coherent states. The total atom number was $400$, with half of these initially in well $a_{1}$ and half in well $b_{2}$.
\subref{fig:number4} gives the expectation values of the population in well $a_{1}$ and \subref{fig:entropy4} gives the calculated pseudoentropies.}
\label{fig:numberentropy4}
\end{figure}

For the four-well system, as shown in Fig.~\ref{fig:numberentropy4}, we see that the maximum pseudoentropy ($\ln 4\approx 1.3863$) is closely approached for both initial Fock and coherent states, but also that the population dynamics are quite different. The small oscillations about the equilibrium value in well $a_{1}$ die down much slower for initial coherent states than for initial Fock states. The off-diagonal elements of the matrix at the end of the integration time were also larger for initial coherent states, although the number distribution is obviously tending towards a quarter of the total in each well.

We note here that for initial Fock states in all these systems, the off-diagonal elements of the reduced density matrix are all zero at $t=0$, while this is not the case for initial coherent states. This can be easy seen by considering some of the typical initial matrix elements for each system. For the two well system with an initial coherent state in one side and nothing in the other, we find
\begin{equation}
\langle\alpha_{1},0|\hat{a}_{1}^{\dag}\hat{a}_{2}|\alpha_{1},0\rangle = \langle\alpha_{1},0|\hat{a}_{2}^{\dag}\hat{a}_{1}|\alpha_{1},0\rangle = 0,
\label{eq:element2C}
\end{equation}
while for a Fock state in one side and nothing in the other, we have
\begin{equation}
\langle N_{1},0|\hat{a}_{1}^{\dag}\hat{a}_{2}|N_{1},0\rangle = \langle N_{1},0|\hat{a}_{2}^{\dag}\hat{a}_{1}|N_{1},0\rangle = 0,
\label{eq:element2F}
\end{equation}
so that all the off-diagonal elements are originally zero. For the three-well system with our coherent state initial conditions, we find two of the off-diagonal elements are non-zero,
\begin{equation}
\langle\alpha_{1},\alpha_{2},0|\hat{a}_{1}^{\dag}\hat{a}_{2}|\alpha_{1},\alpha_{2},0\rangle = \alpha_{1}^{\ast}\alpha_{2} = (\langle\alpha_{1},\alpha_{2},0|\hat{a}_{1}^{\dag}\hat{a}_{2}|\alpha_{1},\alpha_{2},0\rangle)^{\ast},
\label{eq:element3C}
\end{equation}
with expectation values of any operator products containing $\hat{a}_{3}$ or $\hat{a}_{3}^{\dag}$ being zero. For the initial Fock states in wells one and two, we find that all the off-diagonal elements are again initially zero. With the four-well system and the two Fock states, all off-diagonal elements are initially zero, whereas for the two coherent state occupations, we find the non-zero elements,
\begin{equation}
\langle\alpha_{1},0,0,\beta_{2}|\hat{a}_{1}^{\dag}\hat{b}_{2}|\alpha_{1},0,0,\beta_{2}\rangle = \alpha_{1}^{\ast}\beta_{2} = (\langle\alpha_{1},0,0,\beta_{2}|\hat{b}_{2}^{\dag}\hat{a}_{1}|\alpha_{1},0,0,\beta_{2}\rangle)^{\ast}.
\label{eq:element4C}
\end{equation}
We therefore see that, where the initial reduced density matrices are the same for each choice of quantum states, as in the two-well example, we see almost indistinguishable evolution of both the populations and the pseudoentropy. Where they exhibit the most difference, as in the three-well example, the subsequent dynamics are very different. The four-well case, where the initial matrices are less different, lies between these two extremes.  

\section{Conclusions and Discussion}
\label{sec:conclude}

Our stochastic phase-space investigations of the quantum dynamics of these three different Bose-Hubbard models shows that all three may equilibrate for given initial conditions. We have also shown that an effective reduced single-particle density matrix may be calculated using phase-space methods. While this matrix does not have all the information contained in the full quantum density matrices, it does allow for the calculation of a pseudoentropy and an investigation of the equilibriation process. Perhaps more importantly, all the measurements necessary to construct this density matrix are possible in principle. We have shown that classical integrability is not necessarily a good guide as to whether a given system will relax or not, as seen by the two-well system, which does relax to a close to equilibrium state, with only minor revivals over the time we investigated. We have also shown that the presence of initial coherences between the different modes can affect the equilibriation process, as shown by the presence of the off-diagonal elements of the density matrix. We have also demonstrated that the stochastic phase-space methods developed for quantum optics can be useful for the theoretical study of statistical mechanical processes in interacting atomic systems where the Hilbert space becomes so large as to make density matrix methods extremely difficult to use.

What we cannot show, due to the limitations of our numerical methods, is whether the oscillations in these systems ever revive fully in finite time, after their initial collapse. We suffer from two restrictions here. The first is that our exact method, although applicable to the two and three well models, can still not be used for arbitrary times. The second is that the truncated Wigner method gives us accurate information on the collapse of the oscillations, but not on the revivals. However, taken together, these two methods allow us to investigate the medium time dynamics and also allow us to determine the effects of different initial quantum states on the mean-field dynamics. In none of the systems we have examined here do the solutions of the mean-field classical equations even approximate the true dynamics of the mean-fields. In that sense these are truly quantum systems. On a final note, it is interesting to consider the systems we have examined as having non-Markovian reservoirs connected to them at $t=0$, when the unoccupied wells become accessible to the atoms. Whether the relaxation we see can be described accurately as a smaller subsytem coming into some type of equilibrium with this bath will be investigated in later work.

\section*{Acknowledgments}
This research was supported by the Australian Research Council under the Centres of Excellence and Future Fellowships Programs. The authors thank Matthew Davis for stimulating discussions.


\begin{thebibliography}{99}

\bibitem{thermal1}{J.M. Deutsch, \pra {\bf 43}, 2046 (1991).}
%
\bibitem{thermal2}{M. Rigol, V. Dunjko, V. Yurovsky, and M. Olshanii, \prl {\bf 98}, 050405 (2007).}
%
\bibitem{thermal3}{M. Rigol, V. Dunjko and M. Olshanii, Nature {\bf 452}, 854 (2008).}
%
\bibitem{Kinoshita}{T. Kinoshita, T. Wenger, and D.S. Weiss, Nature {\bf 440}, 900 (2006).}
%
\bibitem{zhang}{J.M. Zhang, C. Shen, and W.M. Liu, arXiv:1102.2469v1 (2011).}
%
\bibitem{lfs}{L.F. Santos, A. Polkovnikov, and M. Rigol, \prl {\bf 107}, 040601 (2011).}
%
\bibitem{srednicki}{M. Srednicki, \pre {\bf 50}, 888 (1994).}
%
\bibitem{Berry}{M.V. Berry, \jpa {\bf 10} (1977).}
%
\bibitem{Joel}{G.J. Milburn, J.F. Corney, E.M. Wright, and D.F. Walls, \pra {\bf 55}, 4318 (1997).}
%
\bibitem{Nemoto}{K. Nemoto, C.A. Holmes, G.J. Milburn, and W.J. Munro, \pra {\bf 63}, 013604 (2000).}
%
\bibitem{Anglin}{M.P. Strzys and J.R. Anglin, \pra {\bf 81}, 043616 (2010).}
%
\bibitem{nosso}{C.V. Chianca and M.K. Olsen, \pra {\bf 83}, 043607 (2011).}
%
\bibitem{Wminus1}{R. Graham, Springer Tracts in Modern Physics, {\bf 66}, 1
(1973).}
%
\bibitem{Steel}{M.J. Steel, M.K. Olsen, L.I. Plimak, P.D. Drummond, S.M. Tan, M.J. Collett, D.F. Walls, and R. Graham, \pra {\bf 58}, 4824 (1998).}
%
\bibitem{DanChris}{D.F. Walls and C.T. Tindle, J. Phys. A {\bf 4}, 534 (1972).}
%
\bibitem{QNoise}{C.W. Gardiner, {\em Quantum Noise}, (Springer-Verlag, Berlin, 1991).}
%
\bibitem{nossoEPL}{L.I. Plimak, M.K. Olsen, M. Fleischhauer, and M.J. Collett,
\epl {\bf 56}, 372 (2001).}
%
\bibitem{P+}{P.D. Drummond and C.W. Gardiner, J. Phys. A {\bf 13}, 2353 (1980).}
%
\bibitem{AxMuzza}{M.K. Olsen and A.S. Bradley, \oc {\bf 282}, 3924 (2009).}
%
\bibitem{Eckmann}{J.-P. Eckmann and D. Ruelle, \rmp {\bf 57}, 617 (1985).}
%
\bibitem{Andyhomo}{A.J. Ferris, M.K. Olsen, E.G. Cavalcanti, and M.J. Davis, \pra {\bf 78}, 060104(R) (2008).}
%
\bibitem{Tania}{T.J. Haigh, A.J. Ferris, and M.K. Olsen, \oc {\bf 283}, 3540 (2010).}
%
\bibitem{RCsuper}{M.K. Olsen and L.I. Plimak, \pra {\bf 68}, 031603 (2003).}
%
\bibitem{Wigsuper}{M.K. Olsen, \pra {\bf 69}, 013601 (2004).}
%


\end{thebibliography}
\end{document}